\documentclass[twocolumn,showpacs,preprintnumbers,amsmath,amssymb]{revtex4}
\usepackage{graphicx}% Include figure files
\usepackage{dcolumn}% Align table columns on decimal point
\usepackage{bm}% bold math

\begin{document}

%\preprint{APS/123-QED}

\title{X-ray absorption spectra at the Ca-L$_{2,3}$-edge calculated 
within multi-channel multiple scattering theory}
% Force line breaks with \\

\author{Peter Kr\"uger}
\email{pkruger@u-bourgogne.fr}
\affiliation{
LRRS, UMR 5613 Universit\'e de Bourgogne - CNRS, B.P. 47870, 
21078 Dijon, France
}
\author{Calogero R. Natoli}
\affiliation{
INFN Laboratori Nazionali di Frascati,
Casella Postale 13, I-00044 Frascati, Italy
}

\date{\today}

\begin{abstract}
We report a new theoretical method for X-ray absorption spectroscopy
(XAS) in condensed matter
which is based on the multi-channel multiple scattering theory of 
Natoli {\em et al.}\/ and the eigen-channel R-matrix method.
While the highly flexible  
real-space multiple scattering (RSMS) method guarantees
a precise description of the single-electron part of the problem,
multiplet-like electron correlation effects 
between the photo-electron and localized electrons can
be taken account for in a configuration interaction scheme.
For the case where correlation effects are limited to
the absorber atom, a technique for the solution of the equations
is devised, which requires only little more computation time 
than the normal RSMS method for XAS.
The new method is described and an application to XAS 
at the Ca~$L_{2,3}$-edge in bulk Ca, CaO and CaF$_2$ is presented.
\end{abstract}

\maketitle

\section{Introduction}
Multiple scattering (MS) theory provides an accurate and flexible
scheme for the calculation of unoccupied electronic states
which are probed by various synchrotron experiments such
as X-ray absorption spectroscopy and resonant elastic and in-elastic
X-ray scattering.
The standard theory relies on the single-particle picture,
that is, it neglects electron correlation effects.
This is a great shortcoming, since core-level
X-ray spectra are often strongly modified by electron correlation,
in particular by the Coulomb and exchange interaction
of the valence electrons with the core hole.
In transition metal and rare earth systems, this
interaction can give rise to pronounced atomic multiplet and
satellite structures in the spectra, which can only be accounted
for through many-electron calculations.
A generalization of MS theory to many-electron wave functions was
developed by Natoli {\it et al.}~[1] and is known as
``multi-channel'' MS theory.
Probably the most difficult part of this approach is the 
calculation of the inter-channel potential. 
Here, we propose a reformulation of the theory, 
where the latter problem is completely avoided.
Instead, the multi-channel T-matrix is calculated variationally
using the eigen-channel 
R-matrix method~\cite{kohn48,greene83,lerouzo84,hamacher89,aymar96}.
While R-matrix methods are well known in atomic spectroscopy, 
they have, to our knowledge, never been used for condensed matter
problems.
Michiels {\it et al.}~\cite{michiels97} presented a calculation
of electron energy loss from NiO using an R-matrix method.
They used, however, an atomic model where all solid 
state effects were described phenomenologically using 
an crystal field and a reduced Coulomb interaction.

Here, we present a new formalism for X-ray absorption in condensed
matter, based on multi-channel MS theory and the eigen-channel 
R-matrix method. 
It allows to take account for local electron correlation 
effects in a multi-channel, that is, configuration interaction scheme.
At present, the type of correlations that can be handled on
this level are limited to those between one electron in 
a delocalized state and  
a finite number of electrons/holes in (sufficiently) localized orbitals.
Sufficiently localized means that the wave function is
negligible small beyond the atomic radius.
This applies exactly to inner-core shells and well to the 
$4f$-shell in rare earths. 
Extensions of the method to include correlation effects between
several delocalized electrons are under way.

In this paper, we present the formalism and report results
on the Ca-L$_{2,3}$-edge absorption of different Ca compounds.
The Ca-L$_{2,3}$-edge
is an interesting test case for the new method, 
because the $L_2$ and $L_3$ absorption channels
are strongly coupled through the photo-electron -- core-hole
Coulomb interaction. This leads to a branching ratio of about
1:1, far from the statistical ratio (2:1) which is obtained in 
single-particle theory.
From a point of view of atomic multiplet 
theory~\cite{mansfield76,zaanen85,degroot90}, 
the non-statistical branching ratio is easily 
understood as a case of strong intermediate coupling 
in the ($2p^53d^1$) final state.
The multipole and exchange part of the 2$p$-3$d$ Coulomb interaction
(Slater-integrals $F^k$, $G^k$ with $k>0$) 
is of comparable strength as the 2$p$-spin-orbit interaction,
which gives rise to correlated $2p^53d^1$ final state wave functions, 
where the $2p_{1/2}$ and $2p_{3/2}$-holes are strongly mixed.
The fact that the branching ratio does not change when going from
atomic Ca~\cite{mansfield76} to various Ca compounds~\cite{himpsel91} 
is empirical evidence that this atomic multiplet picture
remains valid in condensed matter.
However, a purely atomic model is not sufficient to account for
fine structure in the $L_{2,3}$-edge spectra,
which depends strongly on the atomic environment~\cite{himpsel91}
(and which is thereby of practical importance for structural 
 and electronic analysis.)
Atomic models including crystal field have proved quite successful
in reproducing the experimental spectra at the 
$L_{2,3}$-edge~\cite{degroot90,himpsel91}.
In that approach, all extra-atomic effects are, however,
treated in an empirical way, by introducing adjustable 
parameters for crystal field and (possibly) band broadening.
Zaanen {\em et al.}~\cite{zaanen85} was the first to 
go beyond the atomic model
by considering a model Hamiltonian that included not only the
atomic $2p$-$3d$ multiplet coupling but also
the single electron density of states of bulk~Ca.
The electron--hole problem was solved exactly using
a Green's function technique. 
While being physically sound, Zaanen's method was not
fully based on first principle calculations, but introduced
a number of empirical parameters.
Later on, Schwitalla and Ebert~\cite{schwitalla98} calculated 
the spectra in the time-dependent local-density-approximation (TD-LDA). 
For bulk Ca, they obtained the correct branching ratio,
but the fine structure of their spectrum was quite different 
from the experimental one.
Recently, Ankudinov {\it et al.}~\cite{ankudinov03} studied
the branching ratio problem with a generalization of TD-LDA.
By adding a frequency and matrix-element dependent 
exchange-correlation contribution to the TD-LDA kernel, 
they obtained a branching ratio in good agreement with 
experiment for Ca and the whole transition 
metal series, while in Ref.~\cite{schwitalla98}
this was true only for the lighter elements (from Ca to~V).
From the theoretical studies cited above, 
it may seem that the
branching ratio problem at the $L_{2,3}$-edge of Ca
has been thoroughly investigated.
Despite of this, we have chosen 
the Ca system as a test case for our new method,
which, we believe, provides new insight into other aspects 
of the problem, like the orbital relaxation around the core-hole
and the reason for the need of a 20\% reduction of the
Slater integrals $F^k$ and $G^k$ in atomic multiplet 
calculations~\cite{degroot90}.
The novelty of the present method comes in at two levels:
it is the first application of the multi-channel MS 
formalism and it is (to our knowledge) the first
true application of $R$-matrix techniques to a 
condensed matter problem.
The combination of these two
features will allow us to shed some light on the two points
mentioned above (orbital relaxation and reduction factors) and
to present an application to the Ca compounds CaO and CaF$_2$,
in which ligand field effects and multiplet structure are
treated in a unique framework in an {\it ab initio}\/ way.

The paper is organized as follows.
In section~\ref{sec_formalism}, the more general aspects of the 
formalism are outlined. Further details about the multi-channel
MS theory can be found in the appendix.
In section~\ref{sec_interaction} the formalism is 
applied to the $L_{2,3}$-edge absorption of $3d^0$ systems
with an emphasis on the screened electron--hole interaction
in the final state.
In section~\ref{sec_numerics} some numerical aspects are discussed.
In section~\ref{sec_results} results are presented for
bulk Ca, CaO and CaF$_2$. 
Finally, some conclusions are drawn in section~\ref{sec_conclusions}.

\section{General formalism}\label{sec_formalism}
In the present approach we go beyond the independent particle 
model by considering a correlated wave function for
a finite number of $N$ electrons.
All other electrons are described within the independent particle 
approximation.
Among the $N$ explicitly treated electrons, at most one is in
a delocalized orbital, all others necessarily occupy localized
orbitals. By definition, a localized orbital is one that is negligibly 
small outside the atomic sphere.
This applies exactly to inner shell orbitals but also to a good 
approximation to 4$f$-orbitals of the rare-earths.
In the ground state wave function, the $N$~electrons include
the core electron that is excited in the XAS process plus
$N-1$ other electrons in localized orbitals.
The XAS final state wave function 
then contains $N-1$ localized electrons 
and one electron (the ``photo-electron'') in a delocalized state above the 
Fermi level.
In other words, we consider a correlated final state wave function 
that couples the photo-electron with the core-hole and 
and/or a finite number of other localized electrons.
%
% The foregoing assumptions consitute the main limitations 
% of the present approach: we go beyond the independant particle
% description only for the interactions among a finite number of electrons, 
% namely for the photo-electron and the core-like electrons.
%

In order to make the derivation less abstract, we shall now
consider the specific case of XAS at the Ca-L$_{2,3}$ edges.
The formulae are kept general and can
easily be applied to other systems to be described with correlated
wave functions satisfying the above requirements.
For the ground state 
we consider the six electron wave function made of the 2$p$ core electrons.
The initial state $\Psi_g$ with energy $E_g$ 
is thus simply given by the closed shell configuration~$(2p^6,^1\!S_0)$.
Final states have energy 
$E = E_g+\hbar\omega$ and a $(2p^5\epsilon^1)$ configuration,
where $\epsilon$ denotes a (one-electron) state in the 
continuum above the Fermi energy. 
The crucial point is that we take into account multiplet 
effects through a configuration interaction ansatz for the
final state wave function, which is developed as 
\begin{equation}\label{eq_psi}
\Psi={\cal A}\sum_{\alpha}{\tilde\Phi}_\alpha(X)\phi_\alpha(x) \;.
\end{equation}
Here ${\tilde\Phi}_\alpha$ is one of the six ($2p^5$) states,
labeled by $\alpha$= $(j_c,\mu_c)$ ($j_c=1/2,3/2$, $\mu_c=-j_c\dots j_c$);
$X$ collects all core-electron coordinates.
The ($2p^5$) multiplet energies are 
$E_\alpha=E_g-\epsilon_c(j_c)$, where $\epsilon_c(j_c)$
are the negative binding energies of spin-orbit split $2p(j_c)$ levels.
For each ${\tilde\Phi}_\alpha$, there is a component $\phi_\alpha$
of the photo-electron wave function.
%, which is developed in 
% a spherical harmonic basis located on the different lattice sites, 
% as usual in multiple scattering theory. 
The (radial, angular, and spin) coordinate of the photo-electron
is  denoted $x=(r,{\hat x},\sigma)$.
Finally, ${\cal A}$ denotes the anti-symmetrization operator.

{\em Multi-channel multiple scattering}.\/
The total photo-absorption cross section is calculated
using the multi-channel multiple scattering method by 
Natoli {\it et al.}~\cite{natoli90}. 
As shown in detail in the appendix, it is given by
\begin{equation}\label{eq_sigma}
\sigma(\omega) \propto \omega {\rm Im}
\left\{ 
   \sum_{\Gamma\Gamma'} M_\Gamma^* \tau_{\Gamma\Gamma'}^{00} M_{\Gamma'} 
\right\} \;.
\end{equation}
Here $\Gamma$=$\alpha Ls$\/ is the set of all quantum numbers of 
$\Psi$, with $L\equiv lm$ being the orbital and $s$ the 
spin quantum numbers of the photo-electron.
$M_\Gamma = \langle \Psi_\Gamma^{\rm in} | D | \Psi_g \rangle$
are the transition matrix elements; we consider only 
dipole transition in the length approximation.
$\Psi_\Gamma^{\rm in}$
is the inside solution that matches smoothly onto the outside 
solution
\begin{equation}\label{eq_psiout}
\Psi_\Gamma^{\rm out} = 
\sum_{\Gamma'}{\Phi}_{\Gamma'}(X{\hat x}\sigma)
Z_{\Gamma'\Gamma}(r)/r \;.
\end{equation}
Here, we have introduced 
${\Phi}_{\Gamma}\equiv{\tilde\Phi}_\alpha(X) Y_{L}({\hat x}) 
\delta_{s,\sigma}$. 
``Inside'' and ``outside'' refer to the atomic sphere
of the absorber, i.e.\ $r<r_0$ and $r>r_0$, 
respectively, $r_0$ being the muffin-tin radius.
The matrix~$Z$ of radial photo-electron functions is given by
\begin{equation}\label{eq_zdef}
Z_{\Gamma\Gamma'}(r)/r=
j_l(k_\alpha r) [t^{-1}_0]_{\Gamma\Gamma'} - i k_\alpha h^+_l(k_\alpha r)
\delta_{\Gamma\Gamma'} \;.
\end{equation}
Here, $h^+_l=j_l+in_l$ and $j_l,n_l$ are the usual spherical Bessel 
and Neumann functions.
$k_\alpha$ is the wave number of the photo-electron,
% in the asymptotic region 
given by $k_\alpha^2 + V_0 = \epsilon_\alpha = E-E_\alpha$, 
where $V_0$ is the interstitial potential.
$t_{\Gamma\Gamma'}$ is the multi-channel atomic T-matrix
of the absorber (at site $i$=0).
In Eq.~(\ref{eq_sigma}), $\tau^{ij}_{\Gamma\Gamma'}$ 
is the multi-channel scattering path operator connecting sites $i$ and~$j$.
It is calculated for a finite cluster by inversion of the
matrix $m\equiv \tau^{-1}$, whose elements are given by 
\begin{equation}\label{eq_mijGG}
m^{ij}_{\Gamma\Gamma'}
=\delta_{ij}[t_i^{-1}]_{\Gamma\Gamma'}
-\delta_{\alpha\alpha'}k_\alpha G^{ij}_{LL'}(k_\alpha)\delta_{ss'} \;.
\end{equation}
Here, $t_i$ is the multi-channel atomic scattering matrix of atom~$i$, and
$G^{ij}_{LL'}$ are the real space KKR structure factors.\cite{structconst}
Apart from the absorber, we treat all atoms in the standard 
one-electron muffin-tin approximation, which implies
$t_{i,\Gamma\Gamma'}=t_{il}(k_\alpha)\delta_{\Gamma\Gamma'}$, for all 
$i$$\ne$0.
Since these T-matrices for $i$$\ne$0 
as well as the structure factors $G^{ij}_{LL'}$ 
are single-channel quantities, 
the only channel-off-diagonal terms of~$m$ are located in $i$=$j$=0 block.
This particular structure of the $m$-matrix allows us to use 
an efficient partitioning technique for the inversion of~$m$.

{\em Partitioning technique.}\/
We divide the system into absorber atom ($i$=0) and ``environment'',
i.e.\ all other atoms with $i$$\ne$0, collectively labeled `$e$'.
For the absorption cross section we need only the absorber block 
$\tau^{00}$ of the $\tau$-matrix.
Using simple matrix algebra, this quantity
can be expressed as
\begin{equation}\label{eq_tau00}
\tau^{00} = \left( 
m^{00} -m^{0e}[m^{ee}]^{-1}m^{e0} 
\right)^{-1}
= \left( 
t_0^{-1} - \rho
% [t_0^{-1}]_{\Gamma\Gamma'} -\rho_{\Gamma\Gamma'}
\right)^{-1} 
\end{equation}
In the second equality, we have used $m^{00} = t_0^{-1}$ and introduced the 
reflectivity $\rho \equiv m^{0e}[m^{ee}]^{-1}m^{e0}$, which
contains all the information we need from the environment.
Once $\rho$ is known, the remaining problem is a purely atomic one.
Now $\rho$ is diagonal in the channel indices~$\alpha$ since it does not
involve the $i$=$j$=0 block of the $m$-matrix.
It can therefore be calculated 
using  standard (single-channel) MS theory. Explicitly, we have
\begin{equation}\label{eq_rhoGG}
\rho_{\Gamma\Gamma'}= 
\delta_{\alpha\alpha'}\rho_{LL'}(k_\alpha)\delta_{ss'} \;,
\end{equation}
where
\begin{equation}
\rho_{LL'}(k) = k^2 \sum G^{0i}_{LL''}(k)
\tilde{\tau}^{ij}_{L''L'''}(k)
G^{j0}_{L'''L'}(k) \;.
\end{equation}
Here, the sum runs over $L''$, $L'''$,
$i$$\ne$0, $j$$\ne$0, and $\tilde{\tau}$ is the single-channel 
$\tau$-matrix of the system without absorber ($\alpha$-diagonal terms
of $[m^{ee}]^{-1}$).

{\it Eigen-channel R-matrix method.}\/
The remaining problem is the calculation of the 
multi-channel T-matrix of the absorber and the inner 
solutions~$\Psi^{\rm in}_\Gamma$. 
This is done using the eigen-channel $R$-matrix 
method~\cite{kohn48,greene83,lerouzo84,hamacher89,aymar96}.
In the following we recall some basic features of this method
for the convenience of the reader and in order to introduce our 
notation (which follows most closely that of Ref.~\cite{hamacher89}).
The R-matrix is a multi-channel generalization of the logarithmic 
derivative of the radial wave function.
As reaction volume, we use the
atomic (or ``muffin-tin'') sphere of the absorbing atom
with radius~$r_0$. 
With Eq.~(\ref{eq_psiout}), the R-matrix can be defined as
\begin{equation}\label{rzdotz}
\sum_{\Gamma''}R_{\Gamma\Gamma''}{\dot Z}_{\Gamma''\Gamma'}(r_0)
=Z_{\Gamma\Gamma'}(r_0) \;.
\end{equation}
Here we have introduced the notation 
${\dot X}\equiv dX/dr$.
Using Eq.~(\ref{eq_zdef}) and its derivative with respect to~$r$,
the $t$-matrix can be readily calculated from the R-matrix as
\begin{equation}\label{eq_tm1}
t^{-1}=iK(R{\dot J}-J)^{-1}(R{\dot H}-H)\;.
\end{equation}
Here all the quantities are matrices with indices ${\Gamma\Gamma'}$
and are evaluated at $r=r_0$.
Furthermore, the quantities $K,J,H$ are diagonal matrices with elements
$K_{\Gamma\Gamma}=k_\alpha$,
$J_{\Gamma\Gamma}=k_\alpha r_0j_l(k_\alpha r_0)$, and
$H_{\Gamma\Gamma}=k_\alpha r_0h^+_l(k_\alpha r_0)$.

In the eigen-channel method, the R-matrix is obtained directly in 
diagonal form; for given energy~$E$, 
a basis of eigenstates $\Psi_k$ 
and eigenvalues ${b_k}$ is found, 
by solving the following generalized eigenvalue 
problem~\cite{hamacher89, aymar96}.
\begin{equation}\label{eq_eigench}
(E-H-L)\Psi_k=Q\Psi_k\,b_k \;.
\end{equation}
Here $H$ is the Hamiltonian,
$L\equiv\sum_{i=1}^{N}\delta(r_i-r_0)
\frac{1}{r_i}\frac{\partial}{\partial r_i}r_i$
is the Bloch operator that restores Hermiticity of~$H$
in the finite reaction volume, i.e.\ the atomic sphere and 
$Q\equiv\sum_{i=1}^{N}\delta(r_i-r_0)$
projects onto its surface.
% The (regular) solutions ($\Psi_k$, $b_k$) are called eigen-channels.
Among all solutions of Eq.~(\ref{eq_eigench}), only those 
with $|b_k|<\infty$ are physically acceptable.
Their number equals the number of channels~$\Gamma$~\cite{lerouzo84}.
In order to solve Eq.~(\ref{eq_eigench}) we develop 
\begin{equation}\label{eq_psik}
\Psi_k = \sum_{\Gamma\nu} \Psi_{\Gamma\nu}c_{\Gamma\nu,k} 
\end{equation}
with trial functions of the form
\begin{equation}\label{eq_psigam}
\Psi_{\Gamma\nu}\equiv{\cal A}
\left\{\Phi_{\Gamma}(X{\hat x}\sigma)P_{\nu}(r)/r\right\} \;.
\end{equation}
As radial basis functions $P_{\nu}$, we use solutions
of the radial Schr\"odinger equation for angular momentum~$l$
and a spherically symmetric, local one-electron potential~$v_{\rm eff}(r)$.
In the present application, we take for~$v_{\rm eff}$ the sum of the ground
state potential~$v_g$ and a partially screened core-hole 
potential~$v_c$ (see Eq.~(\ref{eq_vc}) below).
As usual in the eigen-channel method,
we use closed-type orbitals with boundary conditions $P_{\nu}(r_0)=0$, 
and open-type  orbitals with boundary conditions $dP_{\nu}/dr(r_0)=0$.
Since $2p\rightarrow \epsilon s$ transitions 
have negligible intensity in the near-edge region,
we here include only $l=2$, i.e.\ $d$-waves in the basis.
The generalized eigenvalue problem in Eq.~(\ref{eq_eigench})
is solved using standard numerical routines~\cite{lapack}.
The eigenvectors of the R-matrix are given by 
$W_{\Gamma k}\equiv\sqrt{N}r_0\int \Phi_\Gamma\Psi_k$, where
the integration is over $X{\hat x}\sigma$ and the remaining radial
coordinate of $\Psi_k$ is taken at~$r_0$~\cite{hamacher89}.
The factor $\sqrt{N}$ comes from anti-symmetrization.
From the orthogonality of the channel functions~$\Phi_\Gamma$ 
and Eqs~(\ref{eq_psik},\ref{eq_psigam}) we have 
\[
W_{\Gamma k}= \sum_{\nu}c_{\Gamma\nu,k}P_\nu(r_0)\;.
\]
We normalize the generalized eigenvectors $c_{\Gamma\nu,k}$ ($k$~fixed) 
such that $\sum_{\Gamma} |W_{\Gamma k}|^2 = 1$. Then $W$~is unitarian
and the R-matrix is given by
\[
R_{\Gamma\Gamma'} = 
-\sum_{k}W_{\Gamma k}b_k^{-1}W^\dagger_{k\Gamma'}\;.
\]
The inner solutions that match the outer ones are
\[
\Psi_\Gamma^{\rm in} = \sum_{\Gamma'\Gamma''\nu k}
\Psi_{\Gamma'\nu}c_{\Gamma'\nu,k} 
W^\dagger_{k\Gamma''}Z_{\Gamma''\Gamma}(r_0)\;.
\]

\section{Electron-hole interaction}\label{sec_interaction}
We describe the sub-system of $N$~electrons through 
a Hamiltonian of the form
\[
H^{(N)} = H_0+V = \sum_{i=1}^{N} h_0(i) + V\;.
\]
Here $h_0$ is the one particle Hamiltonian 
of the chosen independent electron model and~$i$ is an electron label.
If $N$ was the number of all electrons in the system ($N_{\rm all}$),
the exact perturbation $V$ would be
given by the bare two-particle electron-electron interaction terms 
minus the effective electron-electron potential $v_{\rm eff}$
that is included in~$h_0$.
However, since in our case $N\ne N_{\rm all}$,
there is no (simple) exact expression of $H^{(N)}$
and the ``best'' approximation for $V$ is not necessarily given by
the exact expression of the case $N=N_{\rm all}$.
The reason is that $v_{\rm eff}$ and thus~$h_0$ are determined by
$N_{\rm all}$ rather than only $N$ electrons, 
and the Coulomb interaction in $V$ is screened 
by the $N_{\rm all}-N$ other electrons.

For the system studied here, these considerations are
of interest only for the final state. The ground state, being 
a closed shell configuration ($2p^6,^1\!S$), is well described by 
a single Slater determinant with the ($2p$) orbitals calculated 
from~$h_0$.
For the $(2p^5\epsilon^1)$ final states, the perturbation~$V$ is
the screened photo-electron -- core-hole Coulomb interaction. 

We shall first take for~$V$ the unscreened interaction
and discuss the effect of screening below.
We have to calculate the matrix elements of $H$, $L$ and $Q$
for the basis states in Eq.~(\ref{eq_psigam}), which we denote as
$|\Gamma\nu\rangle\equiv|2p^5 \nu d^1,\Gamma\rangle$ 
with $\Gamma$=$j_c \mu_c ms$.
% Since by construction, $P_\nu(r)$ is an eigenstate of $h_0(r)$
% with energy~$\epsilon_\nu$,
We have
\[
\langle\Gamma\nu|H_0|\Gamma'\nu'\rangle
=(E_g-\epsilon_c(j_c)+\epsilon_\nu)\delta_{\Gamma\Gamma'}S_{\nu\nu'} \;,
\]
where
\[
S_{\nu\nu'}\equiv \int_0^{r_0}dr P_\nu(r) P_{\nu'}(r)
\]
is the overlap integral, $\epsilon_\nu$ is the energy of the $P_\nu$ orbital, 
and the other quantities have been defined
before. Note that $\delta_{\Gamma\Gamma'}$ is ensured by
the orthogonality of the angular and spin functions.
% but that $S_{\nu\nu'}\ne \delta_{\nu\nu'}$ because open+closed orbs
For the calculation of the matrix elements of~$V$,
we make a basis transformation from the uncoupled states 
$|2p^5 j_c \mu_c, \nu d^1 ms\rangle$ to $LS$~coupled states
$|2p^5 \nu d^1,(LS)JM\rangle$~\cite{basistrafo}.
In the $LS$~coupled basis, the matrix elements of~$V$
are given by 
\begin{equation}\label{eq_VGG}
\langle\Gamma\nu|V|\Gamma'\nu'\rangle =
[w(^{2S+1}L)]_{\nu\nu'} \delta_{\Gamma\Gamma'} \;,
\end{equation}
where now $\Gamma$=$(LS)JM$.
The $w$'s can be expressed in terms of the following
generalized Slater integrals
\[
F^k_{\nu\nu'}\equiv\int_0^{r_0}\!dr\int_0^{r_0}\!dr' 
P_{2p}(r) P_{\nu}(r') \frac{2r^k_<}{r^{k+1}_>}
 P_{2p}(r) P_{\nu'}(r')\;,
\]
\[
G^k_{\nu\nu'}\equiv\int_0^{r_0}\!dr\int_0^{r_0}\!dr'
 P_{2p}(r) P_{\nu}(r') \frac{2r^k_<}{r^{k+1}_>}
 P_{\nu'}(r)P_{2p}(r')\;.
\]
Here, $r_{>(<)}$ is the larger (smaller) of $r$ and $r'$.
The expressions for the $w(^{2S+1}L)$ are given in Ref.~\cite{condon35}:\\
\hspace*{2em}
\begin{tabular}{lll}
$w(^1P)$ &=& $-F^0-F^2/5+4G^1/49$ \\
$w(^3P)$ &=& $-F^0-F^2/5$ \\
$w(^{1,3}D)$ &=& $-F^0+F^2/5$ \\
$w(^1F)$ &=& $-F^0-2F^2/35+18G^3/49$ \\
$w(^3F)$ &=& $-F^0-2F^2/35$ \\
\end{tabular}\\
where, in our case, all quantities are matrices with indices ($\nu\nu'$).
%  another format: in-line
%
%  $w(^1P)=-F^0-F^2/5+4G^1/49$,
%  $w(^3P)=-F^0-F^2/5$,
%  $w(^{1,3}D)=-F^0+F^2/5$,
%  $w(^1F)=-F^0-2F^2/35+18G^3/49$,
%  and $w(^3F)=-F^0-2F^2/35$, 
%
%  yet another format:  in-line with \frac{}{}
%
%  $w(^1P)=-F^0-\frac{1}{5}F^2+\frac{4}{49}G^1$,
%  $w(^3P)=-F^0-\frac{1}{5}F^2$,
%  $w(^{1,3}D)=-F^0+\frac{1}{5}F^2$,
%  $w(^1F)=-F^0-\frac{2}{35}F^2+\frac{18}{49}G^3$,
%  and $w(^3F)=-F^0-\frac{2}{35}F^2$, 
%
The matrix elements of the other operators needed in the eigen-channel
method are easily calculated:
\[
\langle\Gamma\nu|E|\Gamma'\nu'\rangle = E\delta_{\Gamma\Gamma'}S_{\nu\nu'}\;,
\]
\[
\langle\Gamma\nu|Q|\Gamma'\nu'\rangle = 
\delta_{\Gamma\Gamma'}P_{\nu}(r_0){P}_{\nu'}(r_0)\;,
\] 
\[
\langle\Gamma\nu|L|\Gamma'\nu'\rangle = 
\delta_{\Gamma\Gamma'}P_{\nu}(r_0)\frac{P_{\nu'}}{dr}(r_0) \;.
\]
The dipole operator selects $(^1P)$~basis states
and thus only $J=1$ final states give a contribution.
The reduced matrix elements are non-zero for $\Gamma$=($^{2S+1}L_1$,$M$=1)
and given by
\begin{equation}\label{eq_redip}
\langle \Psi_g||r||\Psi_\Gamma^{\rm in}\rangle
 = -2\sum_{\nu k\Gamma'}I_{2p,\nu}
c^{(k)}_{\Gamma_0\nu}W^\dagger_{k\Gamma'}Z_{\Gamma'\Gamma}(r_0)
\end{equation}
with $\Gamma_0$=($^1P_1$,$M$=1) and 
$I_{2p,\nu}=\int_0^{r_0}P_{2p}(r)rP_{\nu}(r)dr$.

{\em Screening model.}\/
As it is well known from cluster and impurity model 
calculations~\cite{zaanen85},
the monopole term of the electron-hole Coulomb 
interaction (corresponding to the Slater integral $F^0$) 
is drastically screened, while the higher order
multipole and all exchange terms (Slater integrals $F^k$, $G^k$ 
with $k>0$) are essentially unscreened.
Let us note that in multiplet calculations 
also the higher order terms are generally reduced from 
the calculated values.\cite{zaanen85,degroot90} 
The need for this reduction of some 20\% is, however, not due to 
screening, but comes mainly from the neglect
of configuration interaction in the single-configuration 
multiplet approach~\cite{cowan80}.
As will become apparent in the next section, the relevant
configuration interaction is included in our approach, so
that there is no need for reduction of the
Slater integrals with~$k>0$.

We therefore apply screening only to the monopole term 
$2/r_>$ of the Coulomb operator~$2/|{\bf x}-{\bf x'}|$.
This defines the (unscreened) multipole part 
$\tilde{V}\equiv 2/|{\bf x}-{\bf x'}|-2/r_>$ of the 
interaction~$V$.
In the space of trial functions~$\Psi_{\Gamma\nu}$ we have chosen,
the operator $2/r_>$ is diagonal in~$\Gamma$.
Within this space, it is therefore equivalent to a one-electron 
potential~$v_u(r)$, namely the Hartree potential of a
spherically symmetric core-hole which is given by~:
$v_u(r)= \int dr' [P_{2p}(r')]^2/r_>$.
We can thus handle screening of the monopole term
on a single-particle level by replacing the unscreened core-hole
potential~$v_u(r)$ by a screened one~$v_c(r)$, which 
we add to~$h_0$. In this way the effective potential $v_{\rm eff}$ used
in $h_0$ will not be the ground state self-consistent potential $v_g$
but $v_{\rm eff} = v_g + v_c$.

A simple approximation for~$v_{\rm eff}$ is given by 
the fully statically screened potential $v_{\rm supercell}$
which is obtained from a self-consistent super-cell calculation with a
core-hole on the absorber site.
This effective potential, which features full orbital 
relaxation around the core-hole,
is frequently used in single-electron XAS calculations.
We shall denote the corresponding core-hole contribution by $v_s$, i.e.\
$v_s \equiv v_{\rm supercell} - v_g$.
As will become apparent in the result section below, 
the line shapes obtained with~$v_s$ are not satisfactory.
We shall therefore allow for incomplete screening by 
using a linear mixture between the unscreened core-hole 
potential~$v_u$ and the fully screened one~$v_s$~:
\begin{equation}\label{eq_vc}
v_c(r) = \alpha v_u(r) + (1-\alpha) v_s(r) \;,
\end{equation}
where $\alpha\in [0,1]$ is an empirical parameter.
As can be seen from the results below, a value of
$\alpha \approx 0.1$ gives best agreement with experiment.
This fact indicates that orbital relaxation around the core-hole is 
overestimated in
$v_{\rm supercell}$ which, we recall, is obtained from a super-cell 
calculation in the {\em local density approximation} (LDA).
This finding was to be expected, since it is known that in LDA the
self-interaction of an electron is not exactly compensated as in the
Hartree-Fock scheme, giving rise to over-relaxation and band gaps that
are systematically too small compared to experiment.
Probably the same calculation with Self-Interaction Corrections 
would cure this drawback. This will be the subject of a future
investigation. In the meantime we regard $\alpha$ as useful parameter
describing the correct amount of relaxation. 
Let us also note that an {\it a priori}\/ estimate of this quantity
could be obtained in a multi-channel MS theory that starts
from the fully relaxed 
state and mixes in very many charge transfer excitations. 
This possibility shall also be explored in the future.

In summary, the present treatment of screening consists in 
(i)~replacing the screened electron--hole Coulomb interaction~$V$ by
its unscreened multipole part $\tilde{V}=2/|{\bf x}-{\bf x'}|-2/r_>$
and (ii)~adding the partially screened core-hole potential~$v_c(r)$ 
in Eq.~(\ref{eq_vc}) to the single-particle Hamiltonian~$h_0$.
Point~(ii) results in a modification of all radial wave
functions~$P_\nu(r)$ and corresponding energies~$\epsilon_\nu$, 
whereas point (i) simply removes all monopole terms ($F^0_{\nu\nu'}$)
from the interaction matrix~(\ref{eq_VGG}).

\section{Numerical aspects}\label{sec_numerics}
The standard MS calculation for the reflectivity of the 
environment~$\rho_{LL'}(k)$ has been performed using the 
CONTINUUM code~\cite{continuum}.
Finite clusters containing at least nine nearest neighbor shells
around the absorber were used for all systems, such that the XAS 
spectra were well converged with respect to cluster size.
The effective single-particle potential was calculated 
self-consistently in the local density approximation 
using the linear-muffin-tin-orbital method~\cite{andersen84}.
In all systems we used space-filling (and thus partially overlapping)
atomic spheres.
In the compounds CaO and CaF$_2$, 
we chose the relative atomic radii in such a way
that the potential value on the sphere was approximately equal for
Ca and the ligand, while keeping the overlap volume small.
For CaO, a ratio of~3:2
between the Ca and~O radii was found appropriate.
For CaF$_2$, the insertion of one empty sphere (E) per formula unit 
was necessary to keep the overlap small. We chose 
the ratio of the sphere radii of Ca:O:E to be 6:4:5, approximately.
The fully screened core-hole potential $v_s$ was obtained from
super-cell calculations with a (spherically 
symmetric) $2p$-hole on the absorber atom.
We used a 32 atom simple cubic super-cell for Ca metal, and an 
fcc $2\times 2\times 2$ super-cell for CaO and CaF$_2$.
We found that the core-hole has only a small 
effect on the potentials of the {\em neighboring}\/ atoms and
that, consequently, it makes hardly a difference for the spectra whether 
the reflectivity is calculated with or without the core-hole potential.
On the absorber atom, however, $v_c(r)$ is strong and
has a dramatic effect on the line shape as will become apparent below.

For reasons of numerical stability, the reflectivity was
calculated at complex energies with a small imaginary part,
such that the spectra are effectively broadened with a 
Lorentzian function of about 0.3~eV FWHM.
In order to simulate finite experimental resolution,
the spectra in the {\em result section}\/ were further broadened with
a Gaussian function of 0.3~eV FWHM.

In the eigen-channel method, convergence 
with respect to the number of radial basis functions has to be achieved.
\begin{figure}
\resizebox{\columnwidth}{!}{\includegraphics{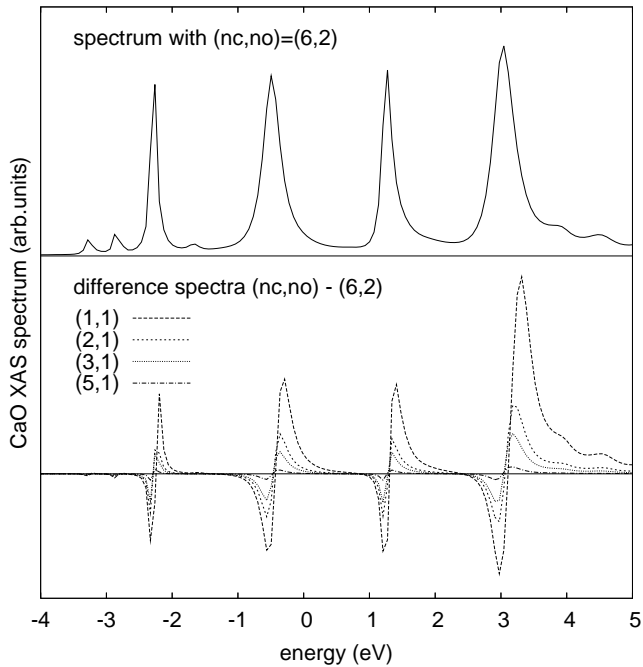}}
\caption{Convergence of the spectra with respect to the number
of radial basis functions in the example of CaO.
The numbers of closed-type ($nc$) and open-type 
($no$) basis functions 
are indicated as ($nc$,$no$).
\label{fig_convergency}
}
\end{figure}
In Fig.~\ref{fig_convergency} we show this convergence
in the example of CaO.
The different basis sets are indicated as ($nc$,$no$), 
where $nc$ ($no$) is the number of closed-type (open-type) functions.
We start from functions without nodes in $0<r<r_0$ and increase the
number of nodes one by one. For example, the (3,1) spectrum was
obtained with three closed type functions of zero, one and two nodes
and one open-type function of zero nodes.
Figure~\ref{fig_convergency} shows
the converged spectrum with basis set (6,2) in the upper panel and 
difference spectra with respect to (6,2) in the lower panel.
It can be seen that five closed-type and only one open-type function
are sufficient for good convergence.
For the spectra in the results section below, we have used
the (5,1) basis set.

It is interesting to note that one can considerably 
improve the spectrum calculated with the minimal (1,1) basis set 
by reducing the values of the Slater integrals $F^k$, $G^k$
(artificially) by some 20\%.
\begin{figure}
\resizebox{\columnwidth}{!}{\includegraphics{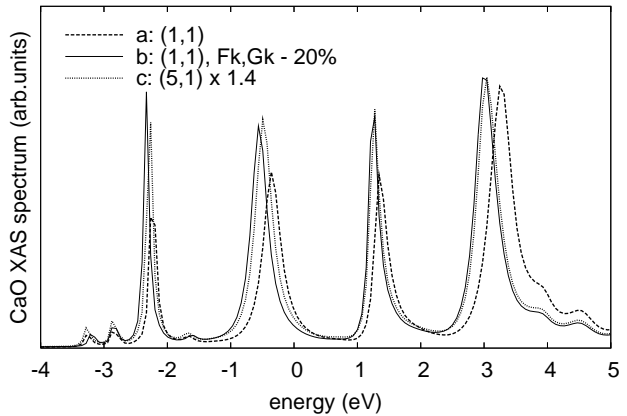}}
\caption{CaO spectra obtained with the minimal basis set ($nc,no$)=(1,1),
along with the converged one (c). The latter was multiplied by a
factor 1.4 for easy comparison.
Spectrum (b) was calculated with the Slater integrals 
$F^k,G^k$ reduced by 20\%.
\label{fig_fkgkrescaled}
}
\end{figure}
Figure~\ref{fig_fkgkrescaled} shows the (1,1) spectrum 
with full (a) and 20\% reduced (b) values of the Slater integrals,
along with the converged spectrum (c), which was multiplied 
by 1.4 for easy comparison of the peak ratios.
Clearly, a 20\% reduction of Slater integrals improves considerably 
the (1,1) line shape, both as far as peak positions and relative peak 
intensities are concerned. Apart from the overall amplitude, 
which is about 40\% too big, the spectrum almost coincides with the
converged one.
This result is closely related to the fact that in 
atomic single configuration multiplet calculations, reduction
factors of 10-25\% for the Slater integrals are generally
needed to make the relative multiplet
energies and line strengths agree with experiment~\cite{cowan80,degroot90}.
Such a rescaling procedure effectively accounts for  
configuration interaction that lies beyond the single configuration 
calculation~\cite{cowan80}, namely coupling to higher lying
electronic configurations.
Precisely this feature is seen in Fig.~\ref{fig_fkgkrescaled},
when one realizes that in the minimal set (1,1) describes essentially 
only the $3d$ orbital, while in the (5,1) basis set of the converged 
spectrum, all $nd$ orbitals up to $n=7$ are included.

In the practical implementation of the method, 
we first calculate the reflectivity matrix $\rho_{LL'}(k)$ 
on a fine mesh in the relevant (photo-electron) energy interval.
In a second step the atomic multichannel calculation is performed
for each total energy $E=E_g+\omega$.
The R-matrix and the inner solutions~$\Psi_\Gamma^{\rm in}$ 
are calculated through the eigen-channel
method and then the atomic multi-channel T-matrix~$t_0$
and the dipole transition matrix elements are readily 
obtained from Eqs~(\ref{eq_tm1}) and (\ref{eq_redip}).
We get the reflectivity $\rho_{LL'}(k_\alpha)$ at the photo-electron
energies $k_\alpha$ of the different channels $\alpha$, 
needed in Eq.~(\ref{eq_rhoGG}) by interpolation in~$k$.
Finally, we invert the matrix $t_0^{-1}-\rho$ (Eq.~\ref{eq_tau00})
and obtain the XAS cross section from Eq.~(\ref{eq_sigma}).

By virtue of the separation between environment and absorber 
through the partitioning technique, the present
implementation of the multi-channel MS method is numerically only
little heavier than the standard (single-channel) MS method.
Indeed, in the present application, 
the atomic multi-channel calculation (second step above) 
was an order of magnitude faster than the reflectivity calculation
by the standard MS technique.

\section{Results for the C\lowercase{a} L$_{2,3}$-edge}\label{sec_results}
\begin{figure}
\resizebox{\columnwidth}{!}{\includegraphics{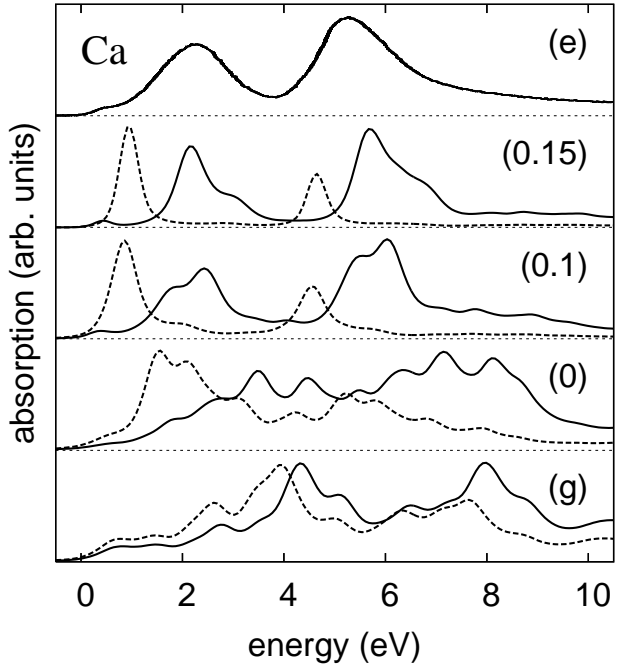}}
\caption{X-ray absorption spectra at the Ca-L$_{2,3}$-edge in bulk Ca.
Curve (e) is the experimental spectrum taken 
from Ref.~\protect\cite{himpsel91}.
In the theoretical spectra, (g) was obtained from the ground state potential,
and the others with a (partially) screened potential with the 
screening factor indicated as ($\alpha$).
In all cases, the one-electron spectrum (without $\tilde{V}$) 
is shown with a dashed line and the multi-channel calculation
(including $\tilde{V}$) with a full line.
\label{fig_caspec}
}
\end{figure}

Figure~\ref{fig_caspec} shows the $L_{2,3}$-edge absorption 
of bulk Ca calculated in different approximations, along with
the experimental spectrum~(e) taken from Ref.~\protect\cite{himpsel91}.
The numbers in parenthesis indicate the value of the screening 
parameter~$\alpha$ of the core-hole potential~$v_c$ in Eq.~(\ref{eq_vc}).
The spectra labeled (g) have been obtained with the 
ground state potential (i.e.\ $v_c=0$).
The spectra in full (dashed) lines have been calculated with (without)
the multipole part of the electron--hole interaction~$\tilde{V}$.
For easy comparison of the line shapes, all spectra are aligned at threshold
and normalized with respect to the height of their main peak.
Note that before normalization, the intensity of the spectra 
without~$\tilde{V}$ (dashed lines) was considerably bigger than 
the corresponding spectra with~$\tilde{V}$ (full lines).
The relative renormalization factors between the two types of spectra,
that have been used in Fig.~\ref{fig_caspec}, are: 
1.8 (g), 2.3 (0), 3.3 (0.1), and 3.8 (0.15).

Probably the most striking feature of the spectra in 
Fig.~\ref{fig_caspec} is the effect of the multipole part of the 
electron--hole interaction~$\tilde{V}$: in all cases,
it leads to a big transfer of spectral
weight from the $L_3$ edge (lower energy peak) to the $L_2$-edge. 
The branching ratio thus changes from $2:1$ without~$\tilde{V}$
to somewhat less than $1:1$, 
which is in good agreement with experiment.
This spectral weight transfer comes from the mixing 
between the $2p_{1/2}$ and the $2p_{3/2}$-hole states
(which correspond, in the one-electron approximation, to 
the $L_2$ and $L_3$ edges, respectively).
It is a genuine atomic multiplet effect
which was first explained by Zaanen {\it et al.}~\cite{zaanen85}.
As can be seen from a ``vertical'' comparison in Fig.~\ref{fig_caspec},
the choice of the core-hole potential~$v_c$
has only a minor effect on the branching ratio, but it 
changes the line shape of the two edges {\it individually}, 
as it can be expected from a single-particle quantity.
Going from (g) to (0), or increasing the parameter~$\alpha$ 
has the effect of shifting the peak positions 
of the two edges to lower energy and of reducing their width.
It moreover leads an overall shift of the whole spectrum to lower energy.
This shift, which is roughly 1~eV for (g)$\rightarrow$(0),
(0)$\rightarrow$(0.1), and (0.1)$\rightarrow$(0.15), is, however,
not apparent from Fig.~\ref{fig_caspec}, because we have aligned the 
spectra at threshold.
When comparing the spectra including~$\tilde{V}$ with the experimental one,
it is clear that (g) and (0) have much too broad peaks.
Moreover, their peak positions relative to threshold are
at too high energy, especially for~(g).
Good agreement for both peak width and positions is obtained for 
spectra (0.1) and (0.15).
The only disagreement is that these two theoretical spectra show a
weak fine-structure which was not observed experimentally.
A possible explanation for this discrepancy is the presence
of further broadening mechanisms,
other than coupling to the band, which is
included here by the multiple scattering of the photo-electron.
Himpsel {\it et al.}\/ suggested that the broadening might be
due to strong auto-ionization.
The discrepancy could, however, also reveal limitations
of the present screening model, which neglects charge fluctuations.
Let us note that our spectrum~(g) looks identical with
the one obtained by Schwitalla {\it et al.}~\cite{schwitalla98}
within time-dependent local density approximation.
This shows that their method does not take account of the
monopole part of the electron--hole interaction.

\begin{figure}
\resizebox{\columnwidth}{!}{\includegraphics{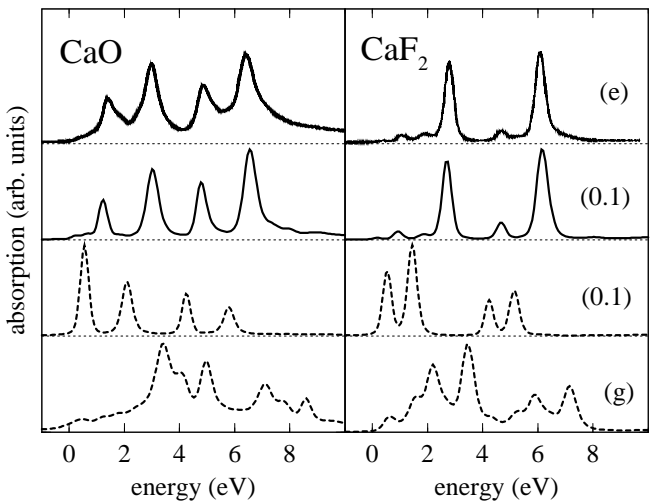}}
\caption{X-ray absorption spectra at the Ca-L$_{2,3}$-edge in Ca0
and CaF$_2$.
Labels and line-styles have the same meaning as in 
Fig.~\protect\ref{fig_caspec},
i.e.\ (e) experiment, 
(g) ground state potential, 
(0.1) $v_c$ with $\alpha$=0.1. 
A full (dashed) line corresponds 
to a calculation with (without)~$\tilde{V}$.
\label{fig_caof2spec}
}
\end{figure}

For a contrast to the {\em metallic}\/ bulk Ca, 
we have applied the method also to two {\em insulating}\/ Ca compounds:
CaO and CaF$_2$.
The results are shown in Fig.~\ref{fig_caof2spec}.
As in Fig.~\ref{fig_caspec}, the spectra have been
normalized and aligned at threshold. For the latter,
the spectra~(g) have been shifted by 3.5~eV to lower energy relative 
to the others in both compounds.
The meaning of labels and line-styles is the same as in Fig.~\ref{fig_caspec}.
The spectra~(g), which correspond to a total neglect of the core-hole
and the multipole terms~$\tilde{V}$,
are again completely at odds with the experimental spectrum~(e).
When using a screened core-hole potential~$v_c$ with $\alpha=0.1$, but 
still neglecting the multipole terms~$\tilde{V}$ [dashed~line~(0.1)],
the spectra consist of four narrow lines (the finite width comes entirely
from the added Lorentzian+Gaussian broadening).
The splitting of the $L_3$ and $L_2$ peaks 
into two doublets is due to a strong ligand field effect.
By symmetry resolved MS calculations, we have checked that for
CaO, the lower (higher) energy peaks correspond to $t_{2g}$ ($e_g$)
symmetry states in the $O_h$ point group.
In CaF$_2$ the order between $t_{2g}$ and $e_g$ peaks is reversed.
These spectra are, however, still very different from the experimental
ones.
When finally also the multipole part of the interaction~$\tilde{V}$ is 
taken into account [full~line~(0.1)], very good agreement with experiment
is obtained for both compounds.
It should be noted that 
Himpsel {\it et al.}~\cite{himpsel91}, who used 
an atomic crystal field model, could also get
very good agreement with experiment.
However, in that work the crystal field
is introduced empirically and its parameter values are adjusted 
to experiment.

\section{Conclusions}\label{sec_conclusions}
In summary, we have presented a method for X-ray absorption 
in condensed matter where single-electron features are 
described in the MS approach, while local multi-electron effects
are taken into account in a configuration interaction scheme.
The novel features of the method are the multi-channel extension of 
MS theory and the use of an R-matrix technique in condensed matter.

The method has been applied to the Ca-L$_{2,3}$ edge absorption
of several Ca systems.
The electron--hole Coulomb interaction was divided into
its monopole and its (higher order) multipole part.
The latter, which is responsible for the non-statistical 
L$_3$:L$_{2}$ branching ratio,
was taken unscreened. We showed that no rescaling for this part is
needed in our method in contrast to single configuration multiplet
calculations.
For the monopole term, a mixture between an
unscreened and a statically screened core-hole potential was
applied.
A mixing factor of about 10\% yields line shapes in good agreement
with experiment in all cases.

Non-local correlation effects such as charge transfer excitations have
been neglected in the present work.
Let us mention, however, that such effects can, in principle, 
be included when the R-matrix reaction volume is extended from a 
single atom to a small cluster of atoms around the absorber.
Compared to recent approaches based on time dependent density functional
theory~\cite{schwitalla98,ankudinov03}, we believe that the present,
configuration interaction based method provides more insight in the
correlation mechanisms at play.
Moreover, the present approach can easily be applied to problems where
the applicability of TD-DFT has yet to be proved, namely 
core-level spectroscopies that involve more than one hole 
(such as Auger processes) or open 4$f$-shells.
Compared to atomic multiplet methods~\cite{degroot90}, 
% (which can handle the latter problems)
the present approach does not rely on adjustable crystal
field parameters. Instead, ligand field and band effects
are described in an {\it ab initio}\/ manner through MS theory.

\section{Acknowledgments}
The authors would like to thank K. Hatada for fruitful discussions.
P. K. acknowledges financial support from the SRRT network and
from INFN.

\appendix
\section*{Appendix}
\subsection*{Derivation of X-ray absorption cross section formula in multi-channel multiple scattering theory}
In this section we derive X-ray absorption cross section formula
within the multi-channel multiple scattering method.
Most of the results of this section are not new, but can
be found as special cases of the more general derivation 
given in Ref.~\cite{natoli90}. 
Nevertheless, we think it is worth including this section
for the convenience of the reader, because the derivation 
is simpler than the one in Ref.~\cite{natoli90} and 
it lends itself better to 
the form of the $N$-electron wave functions used in the
present work.

We start from the general multi-electron formula for
the total optical absorption cross section in the 
dipole approximation~\cite{natoli90}:
\begin{equation}
\label{sigmaf}
\sigma = 4\pi^2\alpha\omega
\sum_{f}\left|\langle \Psi_{f}|D|\Psi_{g}\rangle\right|^2
\delta(E_f-E_g-\omega)
\end{equation}
Here, $\Psi_{g}$ and $\Psi_{f}$ are $N$-electron wave functions,
for initial (=ground) and final state, respectively,
in the absorption process of a photon with energy~$\omega$.
In case of degenerate ground states, a sum over~$g$ is understood.
$D\equiv {\mathbf\epsilon}\cdot\sum_{i=1}^N {\bf x}_i$ is the
dipole operator, $\alpha=\frac{1}{137}$ the fine structure constant.
We use the units: $\hbar=1$,
Bohr radius for length, Rydberg for energy. 
Thus $E_{\rm kin}=k^2$, $e^2=2$.

For the ground state,
we explicitly take into account only localized electrons
of the absorbing atom.
Thus we assume that the ground state wave function $\Psi_g$ 
is confined to the atomic sphere of the absorber
with radius~$r_0$: 
$\Psi_{g}(x_1\dots x_N)=0$ if $\exists i: r_i > r_{0}$.
As for the final state wave function $\Psi_f$, we assume that 
$N-1$~electrons remain in localized orbitals and
at most one electron (the ``photo-electron'') is promoted to a continuum 
orbital.
We chose boundary conditions such that in the remote past, 
the photo-electron is free, i.e.\ its eigenstates are
plane waves $\exp(i{\bf kx})$
times a spin function $\chi_s(\sigma)=\delta_{s\sigma}$.
The rest system is in one of the eigenstates~$\Phi_\alpha(x_1\dots x_{N-1})$
of the $N-1$ electron Hamiltonian with a core hole: 
$H^{N-1}\Phi_{\alpha}=E_{\alpha}\Phi_{\alpha}$.
Thus, the ``incoming part'' of $\Psi_f$  is given by
$\Phi_{\alpha}\times\exp(i{\bf kx})\chi_s(\sigma)$.
In multi-channel scattering theory, not only 
elastic, but also inelastic scattering processes 
are taken into account, which correspond to excitations 
$\Phi_{\alpha}\rightarrow \Phi_{\beta}$.
In the present approach,
these excitations are limited to atomic-like ones, such as
multiplet excitations, due to the local character of $\Phi_{\alpha}$.
(Note that this is in contrast to the more general
theory in Ref.~\cite{natoli90}).
By expanding the scattered part of $\Psi_f$ over
the eigenfunctions $\Phi_\alpha$, we can write
\[
\Psi^{(\alpha{\bf k}s)}=\Phi_{\alpha}\,\exp(i{\bf kx})\chi_s(\sigma)
+\sum_{\beta}\Phi_{\beta}\,f^{(\alpha{\bf k}s)}_{\beta}({\bf x}\sigma)\,.
\]
Here $f^{(\alpha{\bf k}s)}_{\beta}({\bf x}\sigma)$ behaves asymptotically
($r\rightarrow\infty$) like a purely outgoing spherical wave.
In the above form of $\Psi_f$, anti-symmetrization
between the photo-electron and the $N-1$ other electrons
has been disregarded.
We indeed neglect anti-symmetrization for the ``outside solution'',
i.e.\ when the photo-electron is outside the atomic sphere of the
absorber.
For the solution inside the atomic sphere, however, anti-symmetrization 
between all electrons is correctly taken into account
through the eigen-channel method (see main text).
Note that in this work we have, for simplicity, assumed the ``muffin-tin'' 
or more precisely atomic sphere approximation for the 
one-electron potential 
i.e.\ the atomic cells are replaced by space filling spheres with
spherically symmetric potential inside.
The difficulties of multiple-scattering theory arising from 
non-muffin-tin potentials
are essentially independent of the electron correlation problem 
we are dealing with here.
The present multi-channel approach could
easily be generalized to non-muffin-tin multiple scattering methods
in which the muffin-tin spheres are replaced by space-filling atomic cells.
The main change would consist in calculating the R-matrix for 
a sphere surrounding the atomic cell and where the potential in the 
so-called ``moon-region'' (the space outside the cell and inside the sphere)
has been put to zero.

With the final state quantum numbers  $\alpha{\bf k}s$, the
sum in Eq.~(\ref{sigmaf}) becomes 
$\sum_{\alpha s}\int\frac{dk^3}{8\pi^3}$. 
We have $\int\frac{dk^3}{8\pi^3}=\frac{1}{16\pi^3}\int d\hat{k}
\int_0^\infty d\epsilon \sqrt{\epsilon}$, where $\epsilon=k^2$ is
the kinetic energy of the photo-electron. This yields
\[
\sigma = \frac{\alpha\omega}{4\pi}
\sum_{\alpha s}{k}_{\alpha}\int d\hat{k}_{\alpha}
\left|\langle \Psi^{(\alpha{\bf k}_{\alpha}s)}|D|\Psi_{g}
\rangle\right|^2\;,
\]
where ${k}_{\alpha}^2 = E_g + \omega - E_{\alpha}$ from energy conservation.
It is convenient to work in an angular momentum basis,
i.e.\ to use spherical rather than plane waves.
We have
$\int d\hat{k} |{\bf k}\rangle \langle{\bf k}|
= 16\pi^2 \sum_{L} |kL\rangle\langle kL|$,
where $\langle {\bf x}|kL\rangle=j_l(kr)Y_L(\hat{x})\equiv J_L(k{\bf x})$.
Here, $j_l$ are the usual spherical Bessel functions
and $Y_L$ are spherical harmonics.
The cross section now becomes
\begin{equation}
\label{sigmal}
\sigma = {4\pi}{\alpha\omega}
\sum_{\alpha L s}{k}_{\alpha}
\left|\langle \Psi^{(\alpha L s)}|D|\Psi_{g}\rangle\right|^2 \;.
\end{equation}
Here $\Psi^{(\alpha L s)}$ is the scattering state that evolves from the
incoming wave
\begin{equation}
\label{psiinc}
\Psi^{\rm inc}=\Phi_{\alpha}\,J_{L}(k_{\alpha}{\bf x})\chi_{s}(\sigma) \;.
\end{equation}
Following standard multiple scattering theory, we write
the scattered part of the wave as a sum of outgoing
irregular waves from all the centers~i
% $\Psi_i^{\rm sc}$
\[
\Psi=\Psi^{\rm inc}+\sum_i\Psi^{\rm sc}_i
\]
In the following we consider points where the photo-electron coordinate
${\bf x}$ lies outside any muffin-tin sphere (or atomic cell).
For such points the potential is zero and we have
\begin{equation}
\label{psisca}
\Psi^{\rm sc}_i = -i\sum_{\alpha Ls} 
\Phi_\alpha\,k_\alpha H_L(k_\alpha{\bf x}_i)\chi_s(\sigma) B^0_{i\alpha Ls} \;.
\end{equation}
Here ${\bf x}_i\equiv {\bf x}-{\bf R}_i$ and
$H_L(k{\bf x})\equiv h_l^+(kr)Y_L({\hat x})$ where $h_l^+=j_l+in_l$ is 
a Hankel and $n_l$ a spherical Neumann function.
As indicated by the superscript~$0$,
the amplitudes $B^0_{i\alpha Ls}$ depend on the quantum numbers
of $\Psi^{\rm inc}$, which we shall denote ${\alpha_0L_0s_0}$ from now on.
We use the well-known re-expansion theorems:
\begin{equation}
\label{jexpan}
J_L(k{\bf x}_j)=\sum_{L'}J_{L'}(k{\bf x}_i)\Delta^{ij}_{L'L}(k)
\end{equation}
\vspace{-1em}
\begin{equation}
\label{hexpan}
-iH_L(k{\bf x}_j)=\sum_{L'}J_{L'}(k{\bf x}_i)G^{ij}_{L'L}(k)
\end{equation}
where $\Delta^{ij}_{L'L}$ and $G^{ij}_{L'L}$ are the real space 
KKR structure constants.\cite{structconst}
% 
% The former holds always, but the latter only if 
% $|{\bf R}_i-{\bf R}_j|>{\bf x}_i$.
%
Developing 
$\Psi^{\rm inc}$ and the $\Psi^{\rm sc}_j$'s around some given 
center~$i$ and using equations~(\ref{jexpan}) and (\ref{hexpan}), 
respectively, yields:
\begin{equation}
\label{psiab}
\Psi
= \sum_{\alpha Ls}\Phi_\alpha\chi_s\left\{ 
J_{L}(k_\alpha{\bf x}_i) A^0_{i\alpha Ls}
-ik_\alpha H_{L}(k_\alpha{\bf x}_i) 
B^0_{i\alpha Ls}
\right\}
\end{equation}
with
\begin{equation}
\label{ajkhb}
A_{i\alpha Ls}^0
=\delta_{\alpha\alpha_0}\delta_{ss_0} \Delta^{i0}_{LL_0}(k_\alpha)+
k_\alpha\sum_{jL'} G^{ij}_{LL'}(k_\alpha)
B_{j\alpha L's}^{0} \,,
\end{equation}
where the usual convention $G^{ii}_{LL'}\equiv 0$ has been used.

Next we express the exciting wave amplitudes~$A^0_{i\alpha Ls}$ in terms of 
the scattered wave amplitudes~$B^0_{i\alpha Ls}$ at the same site~$i$ through
the inverse atomic scattering matrices~$(t_i^{-1})_{\alpha Ls,\alpha'L's'}$ as 
\begin{equation}
\label{atb}
A^0_{i\alpha Ls}=
\sum_{\alpha'L's'}(t_i^{-1})_{\alpha Ls,\alpha'L's'}B^0_{i\alpha'L's'} \,.
\end{equation}
This holds by definition of the $t_i$-matrices, and relies only on the
most basic assumption of multiple scattering theory, namely that the 
potential can be written as a sum of atomic cell potentials.
Note that there is no restriction on the form of the atomic 
potentials, which may, as it is the case for the absorber potential
in the present work, include non-local and correlation effects.
(Note, however, that in the present approach the 
calculation of this complicated potential is avoided by virtue
of the eigen-channel method.)
Assuming the atomic $t$-matrices to be known, we may use
Eq.~(\ref{atb}) to eliminate the $A^0_{i\alpha Ls}$'s in Eq.~(\ref{ajkhb}),
and then solve for the $B^0_{i\alpha Ls}$'s.
This yields
\begin{equation}
\label{btauj}
B^0_{i\alpha Ls}= \sum_{jL'}
\tau^{ij}_{\alpha Ls,\alpha_0L's_0}
\Delta^{j0}_{L'L_0}(k_{\alpha_0})
\end{equation}
where~$\tau$ is the (multi-channel) scattering path operator which
is defined by its matrix inverse:
\begin{equation}
\label{tauinv}
(\tau^{-1})^{ij}_{\Gamma\Gamma'}\equiv
\delta_{ij}(t_i^{-1})_{\Gamma\Gamma'}
-\delta_{\alpha\alpha'}k_{\alpha}G^{ij}_{LL'}(k_{\alpha})\delta_{ss'}\,,
\end{equation}
where we have introduced the collective index $\Gamma\equiv\alpha Ls$.
Equations~(\ref{psiab}--\ref{tauinv}) 
are the generalized multiple scattering equations.

We shall proceed by calculating the X-ray absorption cross section 
from an atom placed at the origin ${\bf R}_i=0$.
Using Eq.~(\ref{atb}), the wave function in Eq.~(\ref{psiab}) 
around site~$i=0$ (index suppressed) reads
\begin{equation}
\label{psipzb}
\Psi_{\Gamma_0}=
\sum_{\Gamma\Gamma'}\tilde{\Phi}_{\Gamma}
Z_{\Gamma\Gamma'}(r)/r\, B_{\Gamma'}^{\Gamma_0}
\end{equation}
where 
\begin{equation}
\tilde{\Phi}_{\Gamma}\equiv 
{\Phi}_{\alpha}\,Y_L({\hat x})\chi_s(\sigma)
\end{equation}
and
\begin{equation}
\label{zoverr}
Z_{\Gamma\Gamma'}(r)/r\equiv j_l(k_\alpha r) (t^{-1})_{\Gamma\Gamma'}
- i k_\alpha h^+_l(k_\alpha r)\delta_{\Gamma\Gamma'} \,.
\end{equation}
We recall that Eq.~(\ref{psiab}) or (\ref{psipzb}) is valid only in the 
space outside atomic spheres.
For the region inside the atomic sphere of the absorber, we may write  
\begin{equation}
\label{psipb}
\Psi_{\Gamma_0}=
\sum_{\Gamma}{\Psi}^{\rm in}_{\Gamma}B_{\Gamma}^{\Gamma_0}
\end{equation}
where ${\Psi}^{\rm in}_{\Gamma}$ is a solution of the Schr\"odinger
equation inside the atomic sphere, that matches smoothly onto
the outside wave function $\sum_{\Gamma'}\tilde{\Phi}_{\Gamma'}
Z_{\Gamma'\Gamma}(r)/r$.

Putting this into the absorption 
cross section formula, Eq.~(\ref{sigmal}), we obtain
\begin{equation}
\sigma = 4\pi\alpha\omega \sum_{\Gamma\Gamma'\Gamma_0}
\langle\Psi_g|D^\dagger|{\Psi}^{\rm in}_{\Gamma}\rangle
k_{\Gamma_0}B_{\Gamma}^{\Gamma_0}
B_{\Gamma'}^{\Gamma_0*}
\langle{\Psi}_{\Gamma'}^{\rm in}|D|\Psi_g\rangle
\end{equation}
Note that the restriction to $\Psi^{\rm in}$ in the calculation 
of the matrix elements
is valid since we have assumed that $\Psi_g$ vanishes 
outside the atomic cell.
A further simplification of the formula can be achieved if we use
the optical theorem, whose validity in the multi-channel
case was proved in Ref.~\cite{natoli90}:
\begin{equation}
\label{optthe}
\sum_{\Gamma_0} k_{\Gamma_0} 
B_{i\Gamma}^{\Gamma_0}B_{j\Gamma'}^{\Gamma_0*}
=-\frac{1}{2i}\left(\tau-\tau^\dagger\right)^{ij}_{\Gamma\Gamma'}
\end{equation}
If we moreover introduce the notation
$M_{\Gamma}\equiv \langle{\Psi}^{\rm in}_{\Gamma}|D|\Psi_g\rangle$
we finally obtain:
\begin{equation}\label{eq_sigmawithm}
\sigma = -4\pi\alpha\omega\times\Im
\left\{ 
\sum_{\Gamma\Gamma'}
M_{\Gamma}^*\tau^{00}_{\Gamma\Gamma'}M_{\Gamma'}
\right\}
\end{equation}
In this form, the cross section formula
reads exactly as the well-known
one-particle expression (see e.g.\ Ref.~\cite{vvedensky92}).
The fundamental difference is that the quantum numbers $\Gamma$ 
contain internal degrees of freedom of the absorbing atom 
(channels~$\alpha$), which in the present case correspond to
different multi-electron states.


\begin{thebibliography}{9}
\bibitem{structconst}
For a definition of the structure constants, 
see e.g.\ Ref.~\protect\cite{natoli90} appendix~A, 
where, however, they are denoted $H$ and $J$ rather than 
$G$ and $\Delta$, respectively.
\end{thebibliography}
\end{document}